\documentclass{svproc}

\usepackage{url}

\usepackage[numbers]{natbib}
\usepackage{mathtools}
\usepackage{authblk}
\usepackage{amsfonts}
\usepackage{float}
\renewcommand{\refname}{References}  % for the article class
\usepackage[colorlinks = true,
linkcolor = blue,
urlcolor  = blue,
citecolor = blue,
anchorcolor = blue]{hyperref}
\usepackage{subcaption}
\begin{document}
\mainmatter              % start of a contribution
\title{Testing for Network and Spatial Autocorrelation}
\titlerunning{Testing Network Autocorrelation}  % abbreviated title (for running head)
\author{Youjin Lee\inst{1} \and Elizabeth L. Ogburn\inst{2}}
\authorrunning{Youjin Lee and Elizabeth L. Ogburn} % abbreviated author list (for running head)
\tocauthor{Youjin Lee and Elizabeth L. Ogburn}
\institute{University of Pennsylvania, Philadelphia, PA 19104, USA,\\
\and
Johns Hopkins Bloomberg School of Public Health, Baltimore, MD 21211, USA,
\email{eogburn@jhu.edu}.}

\maketitle              % typeset the title of the contribution

\begin{abstract}
Testing for dependence has been a well-established component of spatial statistical analyses for decades. 
In particular, several popular test statistics have desirable properties for testing for the presence of spatial autocorrelation in continuous variables.  In this paper we propose two contributions to the literature on tests for autocorrelation. First, we propose a new test for autocorrelation in categorical variables. While some methods currently exist for assessing spatial autocorrelation in categorical variables, the most popular method is unwieldy, somewhat ad hoc, and fails to provide grounds for a single omnibus test. Second, we discuss the importance of testing for autocorrelation in data sampled from the nodes of a network, motivated by social network applications. 
We demonstrate that our proposed statistic for categorical variables can both be used in the spatial and network setting.
\end{abstract}
\section{Introduction}
\label{sec:intro}

In studies using spatial data, researchers routinely test for spatial dependence before proceeding with statistical analysis~\cite{legendre1993spatial,lichstein2002spatial,diniz2003spatial}.  Spatial dependence is usually assumed to have an autocorrelation structure, whereby pairwise correlations between data points are a function of the geographic distance between the two observations~\citep{cliff1972testing, ord1995local}. 
Because autocorrelation is a violation of the assumption of \emph{independent and identically distributed} (i.i.d.) observations or residuals required by most standard statistical models and hypothesis tests~\citep{legendre1993spatial, anselin1996simple, lennon2000red}, testing for spatial autocorrelation is a necessary step for valid statistical inference using spatial data. 

Taking temporal dependence into account is also widely practiced in time series settings.
But other kinds of statistical dependence are routinely ignored. In many public health and social science studies, observations are collected from individuals who are members of one or a small number of social networks within the target population, often for reasons of convenience or expense. For example, individuals may be sampled from one or a small number of schools, institutions, or online communities, where they may be connected by ties such as being related to one another; being friends, neighbors, acquaintances, or coworkers; or sharing the same teacher or medical provider. If individuals in a sample are related to one another in these ways, they may not furnish independent observations, and yet most statistical analyses in the literature use i.i.d. data methods~\citep{lee2019network}.  This failure to account for dependence can result in anticonservative inference: inflated false positive rates and artificially small p-values.  

In the literature on spatial and temporal dependence, dependence is
often implicitly assumed to be the result of latent traits that are
more similar for observations that are close
than for distant observations. This \emph{latent variable dependence} \citep{ogburn2017challenges} is
likely to be present in many network contexts as well. In networks,
ties often present opportunities to transmit traits or information
from one node to another, and such direct transmission will result
in \emph{dependence due to direct transmission} \citep{ogburn2017challenges} that is informed by the underlying network structure.
In general, both of these sources of dependence result in positive
pairwise correlations that tend to be larger for pairs of observations
from nodes that are close in the network and smaller for observations
from nodes that are distant in the network. Network distance is usually
measured by geodesic distance, which is a count of the number of edges
along the shortest path between two nodes. This is analogous to spatial
and temporal dependence, which are generally thought to be inversely
related to (Euclidean) distance.

Despite increasing interest in and availability of social network
data, there is a dearth of valid statistical methods to account for network
dependence. 
Although many statistical methods exist for dealing with dependent data, almost all of these methods are intended for spatial or temporal data–or, more broadly, for observations with positions in $\mathbb{R}^{k}$ and dependence that is related to Euclidean distance between pairs of points. 
The topology of a network is very different
from that of Euclidean space, and many of the methods that have been
developed to accommodate Euclidean dependence are not appropriate
for network dependence. The most important difference is the distribution
of pairwise distances which, in Euclidean settings, is usually assumed
to skew towards larger distances as the sample grows, with the maximum
distance tending to infinity with sample size $n$. In social networks, on the
other hand, pairwise distances tend to be concentrated on shorter
distances and may be bounded from above. However, as we elaborate
in Section~\ref{sec:method}, methods that have been used to test
for spatial dependence can be adapted and applied to network data.

The most popular tests for spatial autocorrelation use Moran's $I$ statistic~\citep{moran1948interpretation} and Geary's $C$ statistic~\citep{geary1954contiguity} for continuous random variables.  In a companion paper, we show that Moran's $I$ provides valid tests of network dependence whenever the dependence is inversely related to a measure of network distance~\citep{lee2019network}. For categorical random variables, however, available tests based on join count analysis~\citep{cliff1970spatial} are unwieldy and fail to provide a single omnibus test of dependence.
Categorical random variables are especially important in social network settings, where group affiliations are often of interest~\citep{kossinets2006empirical, lewis2008tastes, weaver2018dynamic}.  
Join count analysis has been recently used for testing autocorrelation in categorical outcomes sampled from social network nodes (e.g.~\cite{long2015social}). Farber et al.~\cite{farber2009topology}  proposed a more elegant test for categorical network data and explored its performance in data generated from linear spatial autoregression (SAR) models~\citep{griffith2000linear, lichstein2002spatial}, which are parametric models for network data~\citep{farber2009topology, fujimoto2011network}. As far as we are aware, all of the previous work on testing for network dependence in categorical variables assumes that the data were generated from SAR models, and none of this previous work has considered the performance of autocorrelation tests for more general network settings.  Although SAR models are often used to model network dependent data, there is very little evidence that most social network data truly conform to these models.  In particular, these models cannot capture general forms of latent variable dependence or of dependence due to direct transmission.

In this paper we propose a new test statistic that generalizes Moran's $I$ for categorical random variables. We demonstrate that both Moran's $I$ and our new test for categorical data can be used to test for dependence among observations sampled from a single social network (or a small number of networks).  We assume that any dependence is monotonically inversely related to the pairwise distance between nodes, but otherwise we make no assumptions about the structure of the dependence, and we do not require any parametric assumptions. 
These tests allow researchers to assess the validity of i.i.d statistical methods, and are therefore the first step towards correcting the practice of defaulting to i.i.d.~methods even when data may exhibit network dependence. 

\section{Methods}
\label{sec:method}

\subsection{Moran's $I$}
\label{subsec:Moran}

Moran's $I$ takes as input an $n$-vector of continuous random variables and an $n\times n$ weighted distance matrix $\mathbf{W}$, where entry
$w_{ij}$ is a non-negative, non-increasing function of the Euclidean
distance between observations $i$ and $j$. Moran's $I$ is expected
to be large when pairs of observations with greater $w$ values (i.e.
closer in space) have larger correlations than observations with smaller
$w$ values (i.e. farther in space). The choice of non-increasing
function used to construct $\mathbf{W}$ is informed by background
knowledge about how dependence decays with distance; it affects the
power but not the validity of tests of independence based on Moran's
$I$.

Let $Y$ be a continuous variable of interest and $y_{i}$ be its realized observation
for each of $n$ units~$(i=1,2,\ldots,n)$. Each observation is associated with a location, traditionally in space but we will extend this to networks.  Let $\mathbf{W}$
be a weight matrix signifying closeness between the units, e.g. a matrix of pairwise Euclidean distances for spatial data or 
an adjacency matrix for network data. (The entries $A_{ij}$ in the adjacency matrix $\mathbf{A}$ for a network are indicators of whether nodes $i$ and $j$ share a tie.) Then Moran's $I$ is defined as follows:
\begin{eqnarray}
	I=\frac{\sum\limits _{i=1}^{n}\sum\limits _{j=1}^{n}w_{ij}\big(y_{i}-\bar{y}\big)\big(y_{j}-\bar{y}\big)}{S_{0}\sum\limits _{i=1}^{n}\big(y_{i}-\bar{y}\big)^{2}/n},\label{eq:moran}
\end{eqnarray}
where $S_{0}=\sum\limits _{i=1}^{n}(w_{ij}+w_{ji})/2$ and $\bar{y}=\sum\limits _{i=1}^{n}y_{i}/n$.
Under independence, the pairwise products $(y_{i}-\bar{y})(y_{j}-\bar{y})$
are each expected to be close to zero. On the other hand, under network dependence adjacent
pairs are more likely to have similar values than non-adjacent pairs, and
$(y_{i}-\bar{y})(y_{j}-\bar{y})$ will tend to be relatively large
for the upweighted adjacent pairs; therefore, Moran's $I$ is expected
to be larger in the presence of network dependence than under the
null hypothesis of independence.

\subsection{New methods for categorical random variables}

For a $K$-level categorical random variable, join count statistics compare the
number of adjacent pairs falling into the same category 
to the expected number of such pairs under independence, essentially
performing $K$ separate hypothesis tests. As the number of categories increases, join count analyses become quite cumbersome. Furthermore, they only consider adjacent observations, thereby throwing
away potentially informative pairs of observations that are non-adjacent
but may still exhibit dependence. Finally, the $K$ separate
hypothesis tests required for a join count analysis are non-independent
and it is not entirely clear how to correct for multiple testing. To overcome this last limitation, Farber et al.~\cite{farber2015testing} proposed a single test statistic that combines the $K$ separate joint count statistics.

Instead of extending join count analysis, we propose a new statistic for categorical observations using the logic of Moran's $I$.  This has two advantages over the proposal of \cite{farber2015testing}: it incorporates information from discordant, in addition to concordant, pairs, and it weights pairs according to their probability under the null, allowing more ``surprising" pairs to contribute more information to the test.
To illustrate, under network dependence adjacent nodes are
more likely to have concordant outcomes -- and less likely to have discordant
outcomes -- than they would be under independence. We operationalize
independence as random distribution of the outcome across the network, holding fixed the marginal probabilities of each category. The less
likely a concordant pair (under independence), the more evidence it
provides for network dependence, and the less likely a discordant
pair (under independence), the more evidence it provides against network
dependence. Using this rationale, a test statistic should put higher
weight on more unlikely observations. The following is our proposed
test statistic:

\begin{eqnarray}
	\Phi= \{ \sum\limits _{i=1}^{n}\sum\limits _{j=1}^{n}w_{ij}\big\{2\mathbf{I}(y_{i}=y_{j})-1\big\}/p_{y_{i}}p_{y_{j}} \} / S_{0},\label{eq:phi}
\end{eqnarray}
where $p_{y_{i}}=P(Y=y_{i})$, $p_{y_{j}}=P(Y=y_{j})$,
and $S_{0}=\sum\limits _{i=1}^{n}(w_{ij}+w_{ji})/2$. The term $(2\mathbf{I}(y_{i}=y_{j})-1)\in\{-1,1\}$
allows concordant pairs to provide evidence for dependence and discordant
pairs to provide evidence against dependence. The product of the proportions
$p_{y_{i}}$ and $p_{y_{j}}$ in the denominator ensures that more
unlikely pairs contribute more to the statistic. As the true population
proportion is generally unknown, $\{p_{k}:k=1,...,K\}$ should be
estimated by sample proportions for each category. 

The first and second
moment of $\Phi$ are derived in the Appendix~\ref{ssec:phi}. Asymptotic
normality of the statistic $\Phi$ under the null can also be proven
based on the asymptotic behavior of statistics defined as weighted
sums under some constraints. For more details see Appendix~\ref{ssec:Phiasym}.
For binary observations, which can be viewed as categorical or continuous,
our proposed statistic has the desirable property that the standardized
version of $\Phi$ is equivalent to the standardized Moran's $I$. Tests can be derived based on the asymptotic normal distribution of $\Phi$ under the null, but tests based on the permutation distribution of $\Phi$ when node labels are permuted but the adjacency matrix is held fixed may have better performance in finite sample sizes.

\subsection{Choosing the weight matrix $\mathbf{W}$}

Tests for spatial dependence take Euclidean distances (usually in $\mathbb{R}^{2}$
or $\mathbb{R}^{3}$) as inputs into the weight matrix $\mathbf{W}$. In networks, the entries in {$\mathbf{W}$} can be
comprised of any non-increasing function of geodesic (or other) distance, %for the purposes of the tests for network autocorrelation that we describe below, 
but for robustness we use the adjacency matrix $\mathbf{A}$
for $\mathbf{W}$, where $A_{ij}$ is an indicator of nodes $i$ and
$j$ sharing a tie. The choice of $\mathbf{W}=\mathbf{A}$ puts weight
1 on pairs of observations at a distance of $1$ and weight $0$ otherwise.
In many spatial settings, subject matter expertise can facilitate
informed choices of weights for $\mathbf{W}$ (e.g.~\cite{smouse1999spatial,overmars2003spatial}),
and if researchers have concrete information
about how dependence decays with geodesic network distance then a more informed choice of $\mathbf{W}$ can improve the power of the test.

\section{Simulations}
\label{sec:simulation}

In Section \ref{ssec:catesim}, we demonstrate the validity and performance of our new statistic, $\Phi$, for testing spatial autocorrelation in categorical variables. In Section \ref{ssec:simnetwork}, we demonstrate the performance of $\Phi$ for testing for network dependence.

\subsection{Testing for spatial autocorrelation in categorical variables}
\label{ssec:catesim}

We replicated one of the data generating settings used by Farber et al.~\cite{farber2015testing} and implemented permutation tests of spatial dependence using $\Phi$. First, we generated a binary weight matrix $\mathbf{W}$ with entries $w_{ij}$ indicating whether regions $i$ and $j$ are adjacent. The number of neighbors ($q_{i}$) for each site $i$ was randomly generated through $q_{i} = 1 + \mbox{Binomial}(2(d-1), 0.5)$ for a fixed parameter $d$ that controls the expected number of neighbors.  We simulated $500$ independent replicates of $n=100$ observations under each of four different settings, varying the values of $d=3,5,7,10$. 
We then used $\mathbf{W}$ to generate a continuous, autocorrelated variable:
\begin{align*}
	Y^{*} = (I_{n} - \rho \mathbf{W})^{-1} \mathbf{\epsilon}, \quad \epsilon = \{ \epsilon_{i} \overset{i.i.d.}{\sim} N(0,1):i = 1,\ldots, n\},
\end{align*}
where $I_{n}$ is a $n\times n$ identity matrix and $\rho$ controls the amount of dependence. When $\rho = 0$, $Y^{*}$'s are i.i.d. while positive $\rho$ induces some dependence among $Y^{*}$'s informed by $\mathbf{W}$. Since $Y^{*}$ is continuous, we applied cutoffs based on the $(0.25, 0.5, 0.75)$ quantiles of each simulated dataset to convert $Y^{*}$ into categorical observations $\mathbf{Y} = \left( Y_{1}, Y_{2}, \ldots, Y_{n} \right)$ having $K=4$ categories.

\begin{figure}[ht]
	\centering
	\includegraphics[width=0.7\textwidth]{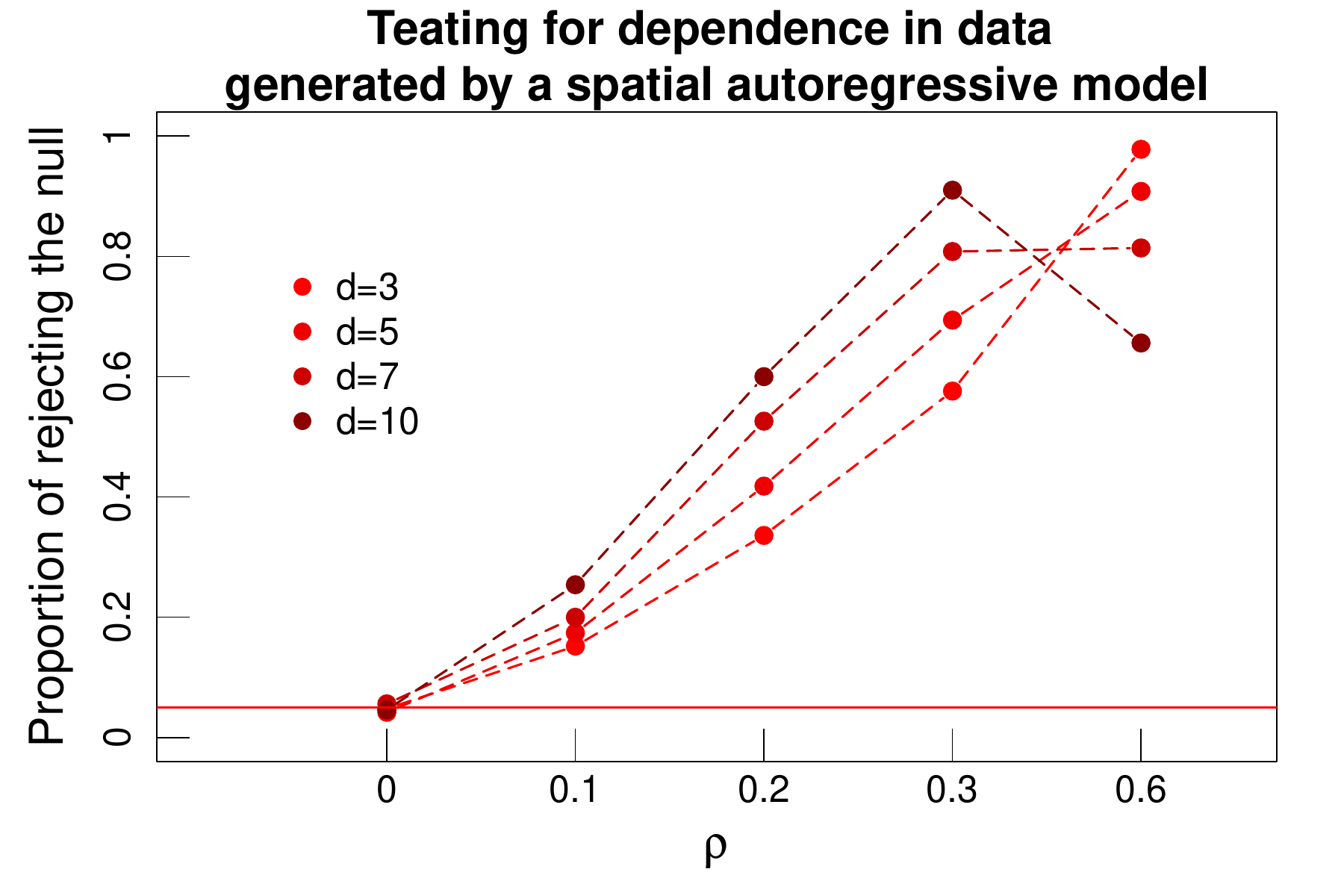}
	\caption{\label{fig:sarsimple} Permutation tests based on $\Phi$. Dependence increases as $\rho$ increases, and the $y$-axis is the proportion of $500$ independent simulations in which the test rejected the null hypothesis of independence.}
\end{figure}

Figure~\ref{fig:sarsimple} presents the simulation results. It shows that under the null ($\rho = 0$), the rejection rate is close to the nominal level of $\alpha = 0.05$ and that the  power to detect dependence increases with $\rho$. 
Moreover, as the expected number of adjacent neighbors, $d$, increases, power tends to increase at fixed $\rho$ when $\rho$ is relatively small ($\rho < 0.6$).  This relationship is reversed when $\rho \geq 0.6$. This can be explained by the fact that, when $\rho$ and $d$ are large relative to $n$, all of the data points tend to look similar to one another, leading to smaller contrasts between pairs of data points that are close and pairs that are distant, i.e. to weaker evidence for dependence.  This is an inevitable feature of any test of dependence that does not rely heavily on a parametric data-generating model.

\subsection{Testing for network dependence}
\label{ssec:simnetwork}

To illustrate the performance of $\Phi$, we simulated categorical
outcomes $Y$ associated with nodes in a single interconnected network and with dependence structure informed by the network ties. $Y$ had five levels and marginal probabilities $(p_{1},~p_{2},~p_{3},~p_{4},~p_{5})=(0.1,~0.2,~0.3,~0.25,~0.15)$; we seeded each node with independent outcomes and then induced dependence due to direct transmission by running a contagious process across the nodes over several time steps; details are provided in the Appendix. The number of time steps, $t$, indexes the amount of dependence induces, with $t=0$ indexing i.i.d. observations. To demonstrate the consequences of using i.i.d. inference in the presence of dependence, as is currently standard practice for network data, we calculated simultaneous 95\% confidence intervals for estimates of $p_{1}$ through $p_{5}$ (using the method proposed in~\citep{sison1995simultaneous}). We tested for network dependence using permutation tests based on $\Phi$ and report power as the percentage of 500 simulations in which the test rejected the null. 

\begin{table}[ht]
	\scriptsize
	\caption{\label{tab:cate_peer_Phi} Coverage rate of simultaneous 95\% CIs, empirical power of tests of independence using asymptotic normality of $\Phi$, and empirical power of permutation tests of independence based on $\Phi$, under direct transmission for $t=0,1,2,3$. The size of the tests is $\alpha=0.05$.}
	\centering
	\resizebox{\textwidth}{!}{\begin{tabular}{|r|r|r|r|}
			\hline
			&  95\% CI coverage rate & \% of p-values($z$) $\leq$ 0.05 & \% of p-values(permutation) $\leq$ 0.05   \\ 
			\hline
			$t$=0 & 0.94 & 5.40 & 4.80 \\ 
			$t$=1 & 0.81 & 39.40 & 36.20 \\ 
			$t$=2 & 0.63 & 67.80 & 65.00 \\ 
			$t$=3 & 0.43 & 85.40 & 83.40 \\ 
			\hline	
	\end{tabular}}
\end{table}

Table~\ref{tab:cate_peer_Phi} summarizes the simulation results. As dependence increases, coverage rates of the 95\% confidence intervals that were estimated under the i.i.d. assumption decrease, representing anticonservative inference.  The power of $\Phi$ to reject the null simultaneously increases. These results indicate (a) that the common practice of using i.i.d. data for network data may be invalid, and (b) that tests based on $\Phi$ can operate as a good screening process for settings in which i.i.d. models are especially problematic. 

The netdep \texttt{R} package for testing network dependence and generating network dependent observations is available through Github (\url{github.com/youjin1207/netdep}).

%%%%%%%%%%%%%%%%%%%%%%%%%%%%%%%%%%%%%%%%%%%%%%%
\section{Applications}
\subsection{Spatial data}
\label{ssec:spatialreal}
In this section we apply $\Phi$ to spatial data on a categorical variable describing the race/ethnicity of populations immediately surrounding  473 U.S. power generating facilities~\cite{papadogeorgou2018adjusting}.  We compare the results to standard analyses using join count statistics.
\begin{figure}[H]
	\centering
	\begin{subfigure}[b]{\textwidth}
		\includegraphics[width=\textwidth]{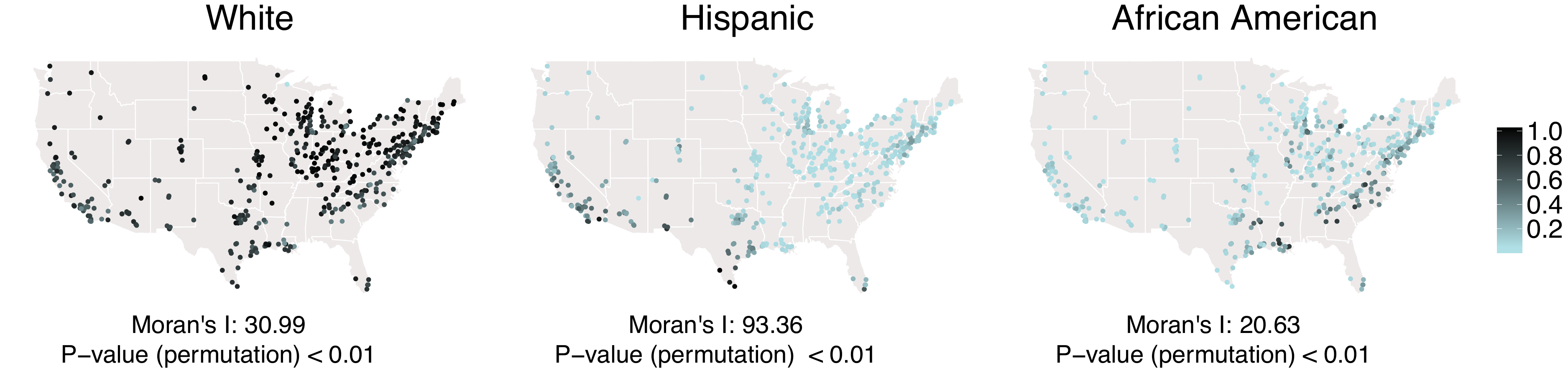}
		\caption{\label{fig:threerace} }
	\end{subfigure}
	\begin{subfigure}[b]{\textwidth}
		\includegraphics[width=\textwidth]{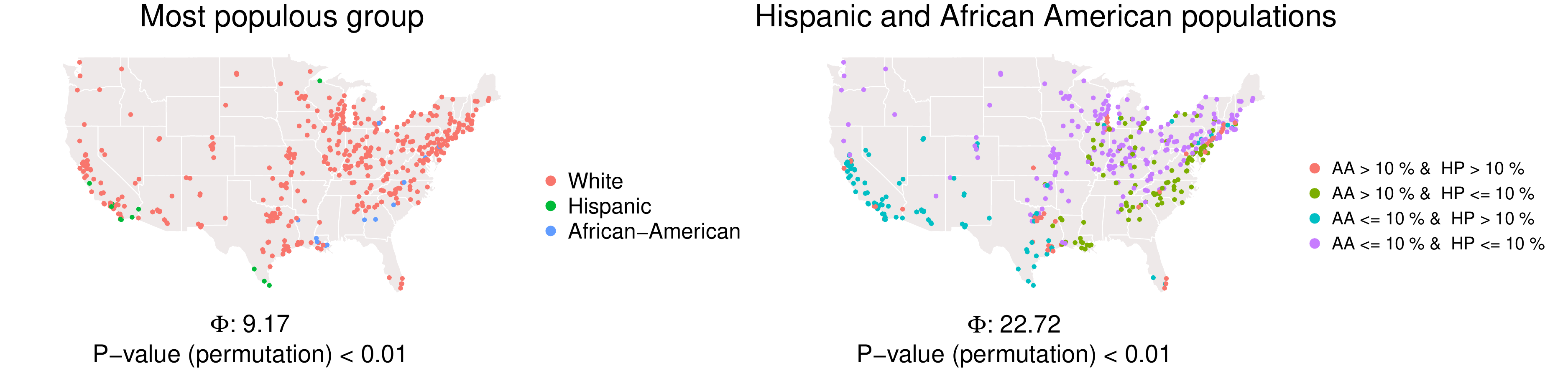}
		\caption{\label{fig:cateplot} }
	\end{subfigure}
	\caption{\label{fig:ethnicity} Panel (a): Proportion of race/ethnicity groups around 473 power-producing facilities across the U.S.. Applying Moran's $I$ separately to each proportion, all of the tests reject the null hypothesis of independence at the $\alpha=0.05$ level.  Panel (b) : Most populous group (left) and categories defined by having $\leq$10\% or $>$10\% Hispanic or African American residents (right). Omnibus tests of dependence based on $\Phi$ reject the null hypothesis of independence at the $\alpha=0.05$ level for both variables.}
\end{figure}

Figure~\ref{fig:threerace} depicts the composition of the population living within a 100~km radius of each power generating facility, with the shade of each dot representing the proportion of the population falling into each race/ethnicity category (White/Hispanic/African American). We can apply Moran's $I$ separately to data on each of the three categories, but Moran's $I$ cannot provide a single aggregate test statistic. Figure~\ref{fig:cateplot} depicts the distributions of two alternative categorical summaries of the information from Figure~\ref{fig:threerace}: a 3-level variable indicating the most populous group in the area surrounding each facility, and a 4-level variable indicating whether more than $10\%$ of the population is Hispanic and African American, respectively.  Using each of these categorial variables, we can perform an omnibus test for dependence using $\Phi$. We observe greater evidence of dependence in the second categorization ($\Phi : 22.72$) than the first categorization ($\Phi : 9.17$).  
%\yl{[are the p-values really identical for every test in this figure?? And in the two tables?????]} 
This direct comparison is possible using $\Phi$ but would not be possible using join count statistics. The join count statistics for these two categorical variables are given in Table~\ref{tab:dominant} and Table~\ref{tab:nonwhite}.  
%For example, we may be interested in autocorrelation with respect to the dominant demographic group (Table~\ref{tab:dominant}) or regions with more than $10\%$ Hispanics and African Americans (Table~\ref{tab:nonwhite}). 
The statistics themselves count the frequency of concordant neighboring pairs for each category and standardize it;
%\yl{[it isn't a count exactly, what is it? it's ok for it to be less than 1?]}, 
 the p-values are derived from a permutation test that permutes the location of each observation while holding the values fixed.  Join count analysis requires a notion of adjacency; we specified a neighborhood size of 15, meaning that observation $j$ is considered to be adjacent to $i$ if $j$ is one of $i$'s closet 15 neighbors in Euclidean distance.

\begin{table}[ht]
	\centering
	\caption{\label{tab:dominant} Permutation tests of dependence based on join count statistics applied to the most populous group.}
	\resizebox{0.6\textwidth}{!}{\begin{tabular}{|l|l|l|l|}
			\hline
			Most populous group & White & Hispanic & African-American \\ 
			\hline
			$n$ & 446 & 13 & 14 \\
			Join count statistic & 212.63 & 0.97 & 0.77 \\ 
			P-value (permutation) & $< 0.01$ & $< 0.01$ & $< 0.01$ \\ 
			\hline
	\end{tabular}}
\end{table}
\begin{table}[ht]
	\centering
	\caption{\label{tab:nonwhite} Permutation tests of dependence based on join count statistics applied to four different population categories, defined by having $\leq$10\% or $>$10\% Hispanic or African American residents. }
	\resizebox{1.0\textwidth}{!}{\begin{tabular}{|r|l|l|l|l|}
			\hline
			& AA $>$ 10\%, HP $>$ 10\% &  AA $>$ 10\%, HP $\leq$ 10\% &  AA $\leq$ 10\%, HP $>$ 10\% &
			AA $\leq$ 10\%, HP $\leq$ 10\% \\ 
			\hline
			$n$ & 52 & 106 & 98 & 217 \\
			Join-count statistic & 7.07 & 26.63 & 30.30 & 69.20 \\ 
			P-value (permutation) & $< 0.01$ & $< 0.01$ & $< 0.01$ & $< 0.01$ \\ 
			\hline
	\end{tabular}}
\end{table}

\subsection{Network data}
\label{ssec:autocorrelation}

The Framingham Heart Study, initiated in 1948, is an ongoing cohort
study of participants from the town of Framingham, Massachusetts that
was originally designed to identify risk factors for cardiovascular
disease. The study has grown over the years to include five cohorts. 
For decades, FHS has been one of the most successful and influential
epidemiologic cohort studies in existence. It is arguably the most
important source of data on cardiovascular epidemiology. It has been
analyzed using i.i.d. statistical models (as is standard practice
for cohort studies) in over 3,400 peer-reviewed publications since
1950: to study cardiovascular disease etiology (e.g.~\cite{castelli1988cholesterol,d2008general}),
risks for developing obesity (e.g.~\cite{vasan2005estimated}),
factors affecting mental health (e.g.~\cite{qiu2010physical,saczynski2010depressive}), and many
other outcomes.

In addition to being a very prominent cohort study, more recently FHS has played
a uniquely influential role in the study of social networks and social
contagion. Researchers reconstructed the (partial) social network underlying the cohort and used this network to study social contagion and peer influence for a variety of outcomes in a series of highly influential papers~\citep{christakis2007spread, christakis2008collective, fowler2008dynamic}.  However, even these analyses use methods that assume independence across subjects~\cite{lyons2011spread,lee2019network}.  In a companion paper we test for dependence in continuous and binary variables in the FHS data, and discuss the implications of network dependence for the body of research that relies on i.i.d. analyses these data.  Here we illustrate that dependence in these data may extend beyond continuous and binary variables to categorical variables, which previous methods would not have been able to ascertain. 
We analyzed $n=$1,033 subjects with 690 undirected social network ties from the Offspring Cohort at Exam 5, which was conducted between 1991 and 1995.

\begin{figure}[H]
	\centering
	\begin{subfigure}[b]{0.4\textwidth}
		\includegraphics[width=\textwidth]{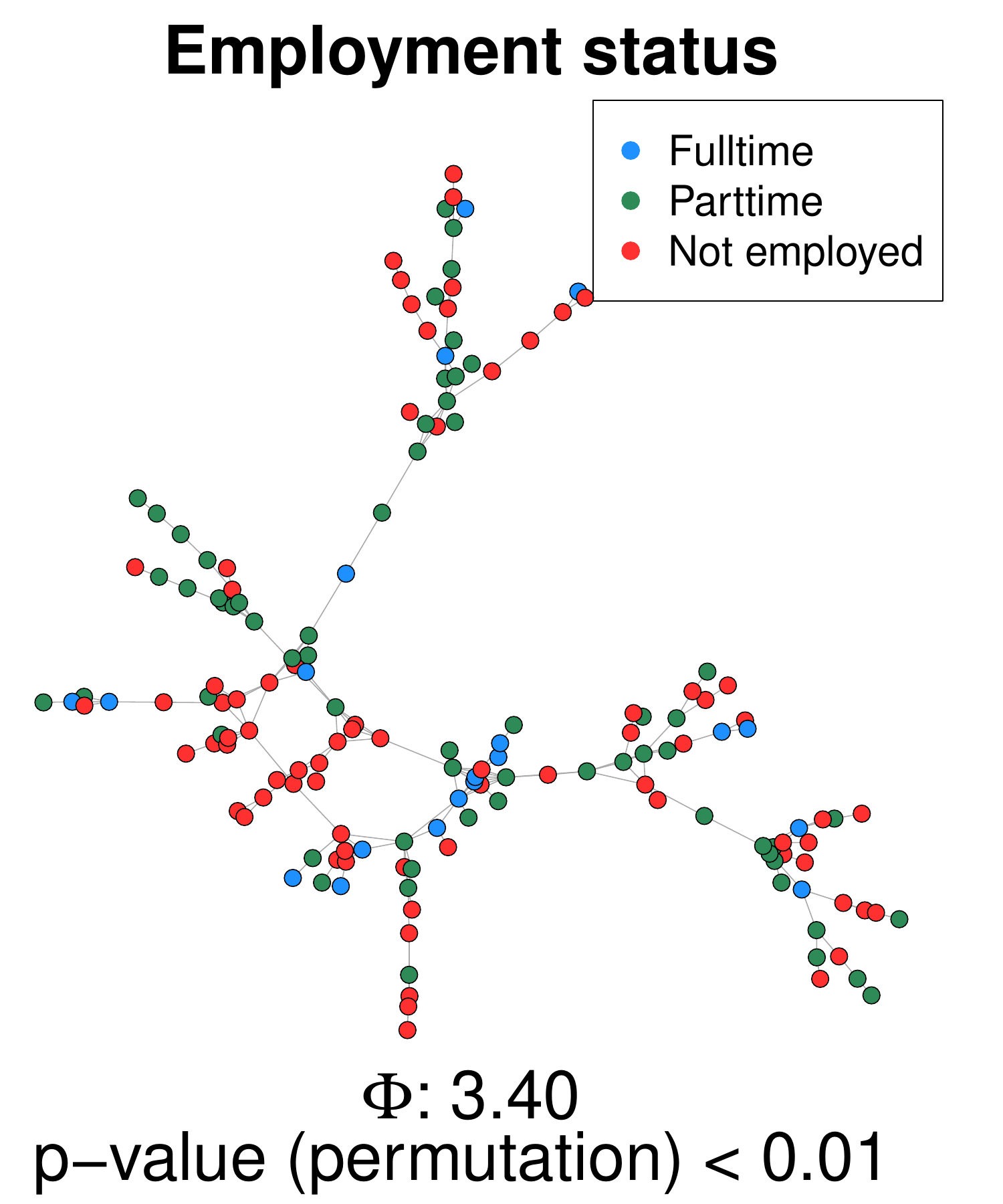}
	\end{subfigure}	\hspace*{2cm} 
	\begin{subfigure}[b]{0.4\textwidth}
		\includegraphics[width=\textwidth]{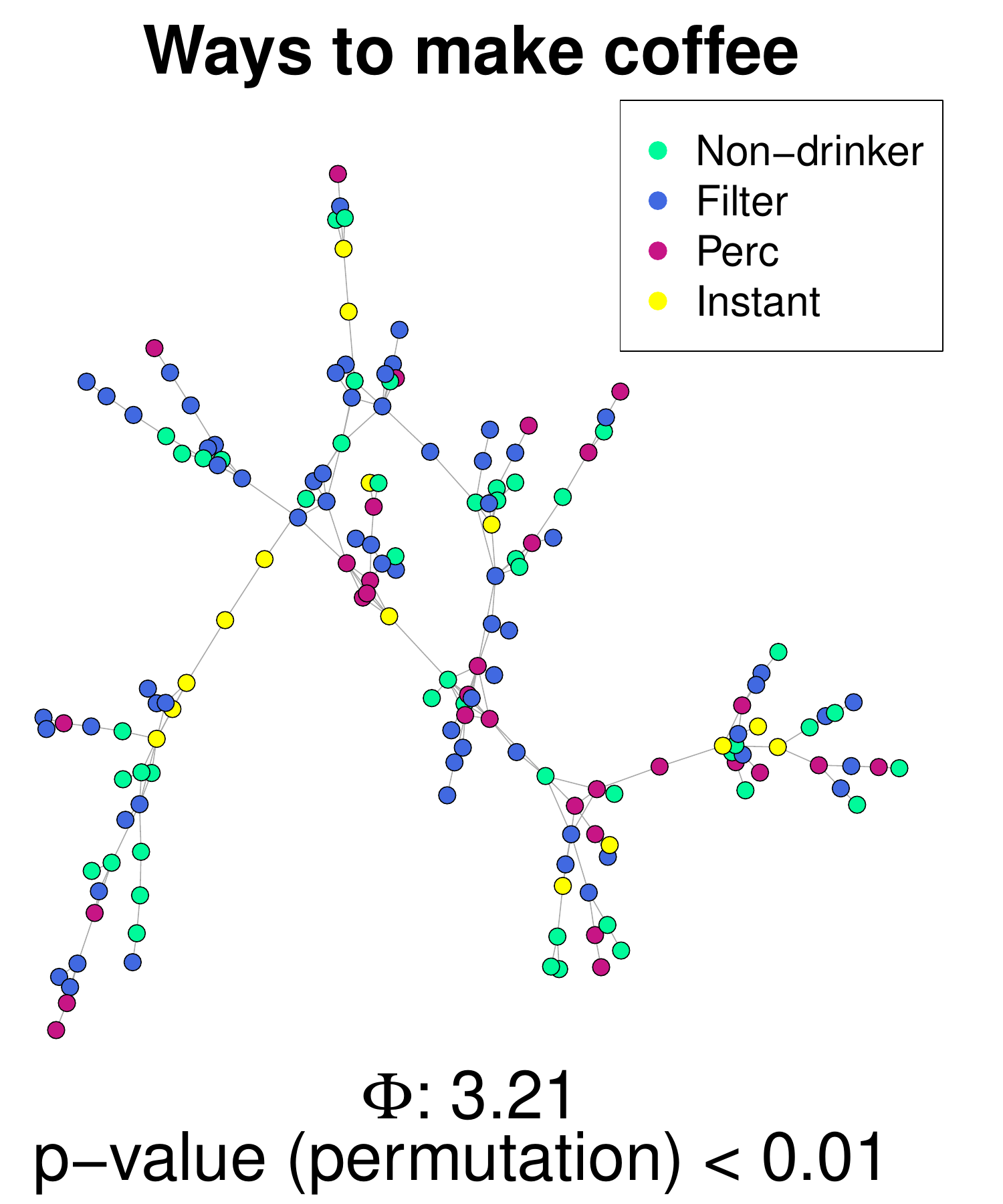}
	\end{subfigure}
	\caption{\label{fig:FHS_cate} Network dependence test for categorical variables with three levels (left) and four levels (right) using $\Phi$.}
\end{figure}

We tested for dependence in two different categorical random variables using $\Phi$: employment status and preferred method of making coffee.  Figure~\ref{fig:FHS_cate} shows the distribution of these two variables over the largest connected component of the network. We found significant evidence of network dependence for both variables, resulting p-value of $<0.01$ in both variables. 

%%%%%%%%%%%%%%%%%%%%%%%%%%%%%%%%%%%%%%%%%%%%
\section{Concluding Remarks}
\label{sec:conclusion}

In this paper, we proposed a simple test for dependence among categorical observations sampled from geographic space or from a network. We demonstrated the performance of our proposed test in simulations under both spatial and network dependence, and applied it to spatial data on U.S. power producing facilities and to social network data from the Framingham Heart Study. 

Under network dependence, adjacent pairs are expected to exhibit the greatest correlations, and for robustness we used the adjacency matrix as the weight matrix for calculating the test statistic, thereby restricting our analysis to adjacent pairs; if researchers have substantive knowledge of the dependence mechanism other weights may increase power and efficiency.

Researchers should be aware of the possibility of dependence in their observations, both when studying social networks explicitly and when observations are sampled from a single community for reasons of convenience. As we have seen in the classic Framingham Heart Study example, such observations can be dependence, potentially rendering i.i.d. statistical methods invalid. In a companion paper~\citep{lee2019network}, we delve deeper into the consequences of assuming that observations are independent when they may in fact exhibit network dependence.  That paper focuses on continuous and binary variables, but similar conclusions hold for the categorical variables that we addressed in this paper.

\section*{Acknowledgments}
\addcontentsline{toc}{section}{Acknowledgment}

Youjin Lee and Elizabeth Ogburn were supported by ONR grant N000141512343. The Framingham Heart Study is conducted and supported by the National
Heart, Lung, and Blood Institute (NHLBI) in collaboration with Boston
University (Contract No. N01-HC-25195 and HHSN268201500001I). This
manuscript was not prepared in collaboration with investigators of
the Framingham Heart Study and does not necessarily reflect the opinions
or views of the Framingham Heart Study, Boston University, or NHLBI.

\appendix
\section{Appendix}
\label{sec:appendix}

\subsection{Moments of $\Phi$}
\label{ssec:phi}

Here we derive $\mu_{\Phi} : = E [ \Phi ]$ and $E[  \Phi^2 ]$, the first and second moments of $\Phi$. Based on these moments, we can derive the variance of $\Phi$, $\sigma^{2}_{\Phi} := E[\Phi^2] - \mu^2_{\Phi}.$  When $K$ is the number of categories and $p_{j}$ is the proportion of $Y$ in category $j~(j=1,2,\ldots, K)$,
\begin{equation}
\begin{split}
\label{eq:phim2}
\mu_{\Phi}  & = \frac{1}{n(n-1)} \{n^2 K (2-k) - n Q_{1} \}, \\
E[\Phi^2] & = \frac{1}{S^{2}_0}  \Bigg[  \frac{S_{1}}{n(n-1)} (n^2 Q_{22} - n Q_{3}) \\  & +  \frac{S_{2} - 2 S_{1}}{n(n-1)(n-2)} ((K-4) K+4) n^3 Q_{1}+n (n ((2 K-4) Q_{2}-Q_{22})+2 Q_{3})  \\ & +  \frac{S^2_{0} - S_{2} + S_{1} }{n(n-1)(n-2)(n-3)}  \Big\{ n (-4 Q_{3} + 2 n Q_{22} - 6 K n Q_{2} + 12 n Q_{2} \\ &-  3 K^2 n^2 Q_{1} + 14 K n^2 Q_{1} - 16 n^2 Q_{1} + K^4 n^3 - 4 K^3 n^3 + 4 K^2 n^3) \\  & -  ((2 K - 4) n^2 Q_{2} + n^2 (K n (2 Q_{1} - K Q_{1}) - Q_{22}) + 2 n Q_{3}) \Big\}    \Bigg],
\end{split}
\end{equation}
where $Q_{m} : = \sum\limits_{l=1}^{K} 1 / p^{m}_{l}, (m=1,2,3)$; $Q_{22} : = \sum\limits_{l=1}^{K} \sum\limits_{u = 1}^{K} 1/p_{l} p_{u}$ ; $S_0 = \sum\limits_{i=1}^n \sum\limits_{j=1}^n (w_{ij} + w_{ji})/2$; $S_1 = \sum\limits_{i=1}^n \sum\limits_{j=1}^n (w_{ij} + w_{ji})^2/2$; $S_2 = \sum\limits_{i=1}^n (w_{i\cdot} + w_{\cdot i})^2$.

\subsection{Asymptotic distribution of $\Phi$ under the null}
\label{ssec:Phiasym}

Shapiro and Hubert~\citep{shapiro1979asymptotic} proved the asymptotic normality of  permutation statistics of the form $H_{n}$ for i.i.d. random variables $Y_{1}, Y_{2}, \ldots, Y_{n}$ under some conditions: 
\begin{equation}
H_{n} = \sum\limits_{i=1}^{n} \sum\limits_{j=1, j \neq i}^{n} d_{ij} h(Y_{i} , Y_{j}),
\end{equation} 
where $h(\cdot, \cdot)$ is a symmetric real valued function with $E[h^2 (Y_{i}, Y_{j})] < \infty$ and $\mathbf{D} : = \{d_{ij} ; i,j=1,...,n \}$ is a  $n \times n$ symmetric, nonzero matrix of which all diagonal terms must be zero. In the context of $\Phi$, $h (Y_{i}, Y_{j}) = \big( 2 I (Y_{i} = Y_{j}) - 1 \big) / (p_{Y_{i}} p_{Y_{j}})$ and $\mathbf{D} = \mathbf{W}.$ 
Requirements for asymptotic normality include $\sum\limits_{i,j=1, j \neq i}^{n} d^2_{ij} / \sum\limits_{i=1}^{n} d^2_{i \cdot} \rightarrow 0$ and $\max\limits_{1 \leq i \leq n}d^2_{i \cdot} / \sum\limits_{k=1}^{n} d^2_{k \cdot} \rightarrow 0$ as $n \rightarrow 0$ for $d_{i \cdot} = \sum\limits_{j=1}^{n} d_{ij}$. If we use the adjacency matrix for $\mathbf{W}$, this implies $\sum\limits_{i,j = 1, i \neq j}^{n} A_{ij} / \sum\limits_{i=1}^{n} A^2_{i \cdot} \rightarrow 0$ and $\max\limits_{1 \leq i \leq n} A_{i \cdot} / \sum\limits_{i=1}^{n} A^2_{i\cdot} \rightarrow 0$ where $A_{i \cdot}$ is the degree of node $i$. More details can be found in~\cite{shapiro1979asymptotic}; see also \cite{o1993asymptotic}.

\section{Simulation of categorical observations over network}
\label{ssec:cate}

\subsection{Direct transmission simulations}	

We specify the starting probability that each observation falls into one of $K$ categories, $\{(p_{1}, p_{2} , ... , p_{K}) : \sum\limits_{j=1}^{K} p_{k} = 1  \}$. We then simulate initial outcomes from a multinomial distribution, and generate outcomes at subsequent time points iteratively:

\begin{equation}
\begin{split}
& Y^{0}_1, Y^{0}_2, \ldots , Y^{0}_{n}  \overset{i.i.d}{\sim} \mbox{Multinomial}\big( (p_1, p_2, \ldots , p_K) \big) \\
& Y^{t}_i = \left\{ \begin{array}{ll} Z_{i}^{t} \sim \mbox{Multinomial}\big( (\hat{p}^{t-1}_{i1}, \hat{p}^{t-1}_{i2}, \ldots , \hat{p}^{t-1}_{iK}) \big) &  \mbox{ with probability } q \\ Y^{t-1}_{i} & \mbox{ with probability } 1 - q  \end{array} \right. 
\end{split}
\end{equation}
where $\hat{p}^{t-1}_{im} : = \sum\limits_{j=1}^{n} w_{ij} I(y^{t-1}_{j} = m ) / \sum\limits_{j=1}^{n} w_{ij};~m = 1, .. , K;~0 < q \leq 1.$
At each time point, with probability $q$, a node's outcome is updated as a draw from a new multinomial with probabilities influenced by the proportion of adjacent nodes falling into each category at the previous time. The amount of influence from adjacent peers can be controlled by pre-specified maximum susceptibility probability $q_{m} (0 \leq q_{m} \leq 1),$ where  $q \in [0, q_{m} ],$ and we set $q_{m} = 0.4$.

\bibliographystyle{spbasic}
\bibliography{reference}


\end{document}